\documentclass[fleqn,usenatbib,letters]{mnras}

\usepackage{comment}

\usepackage[T1]{fontenc}

\DeclareRobustCommand{\VAN}[3]{#2}
\let\VANthebibliography\thebibliography
\def\thebibliography{\DeclareRobustCommand{\VAN}[3]{##3}\VANthebibliography}

\defcitealias{}{}

\usepackage{graphicx}	
\usepackage{amsmath}	

\usepackage[dvipsnames]{xcolor}
\usepackage{txfonts}

\usepackage{mathrsfs}

\usepackage[dvipsnames]{xcolor}

\title{Gravitationally Lensed Orphan Afterglows of Gamma-Ray Bursts}
\author[]{
Hao-Xuan Gao,$^{1}$
Jin-Jun Geng,$^{2}$ \thanks{E-mail: jjgeng@pmo.ac.cn}
Lei Hu,$^{2}$
Mao-Kai Hu,$^{2}$
Guang-Xuan Lan,$^{2,3}$
Chen-Ming Chang,$^{2,3}$
\newauthor
Song-Bo Zhang,$^{2}$
Xiao-Li Zhang,$^{1}$
Yong-Feng Huang,$^{4,5}$ \thanks{E-mail: hyf@nju.edu.cn}
Xue-Feng Wu$^{2,3}$ \thanks{E-mail: xfwu@pmo.ac.cn}
\\
$^{1}$School of Physics, Nanjing University, Nanjing 210023, People's Republic of China \\
$^{2}$Purple Mountain Observatory, Chinese Academy of Sciences, Nanjing 210023, People's Republic of China \\
$^{3}$School of Astronomy and Space Sciences, University of Science and Technology of China, Hefei 230026, China \\
$^{4}$School of Astronomy and Space Science, Nanjing University, Nanjing 210023, People's Republic of China\\
$^{5}$Key Laboratory of Modern Astronomy and Astrophysics (Nanjing University), Ministry of Education, Nanjing 210023, People's Republic of China\\
}

\date{Accepted XXX. Received YYY; in original form ZZZ}

\pubyear{2022}

\begin{document}
\label{firstpage}
\pagerange{\pageref{firstpage}--\pageref{lastpage}}
\maketitle

\begin{abstract}
  The cosmological nature of gamma-ray bursts (GRBs) implies that a small portion of them could
  be gravitationally lensed by foreground objects during their propagation.
  The gravitational lensing effect on the GRB prompt emission and on-axis afterglows has been discussed,
  and some candidates have been found in the literature.
  In this work, considering the high detection rate of GRB orphan afterglows in future wide-field survey era,
  we investigate the gravitationally lensed orphan afterglows in view of three lens models, i.e., the point-mass model, the singular isothermal sphere model,
  and the Chang-Refsdal model. The structure of the GRB jet itself is also incorporated in calculating the lensed afterglow light curves.
  It is found that lensed optical/X-ray orphan afterglows in principle could be diagnosed through their temporal characteristics, and the optical band is the best band to
  observe the galaxy-lensed orphan afterglows.
  Moreover, the event rate for galaxy-lensed orphan afterglows is estimated to be  $\lesssim$ 1.8 $\text{yr}^{-1}$ for the whole sky.
  If most orphan afterglows could be identified (from other transients in the survey data),
  the optimistic detection rates of the 2.5m Wide Field Survey Telescope of China and 8.4m Vera Rubin Observatory Legacy Survey of Space and Time
  for galaxy-lensed orphan afterglows in the optical band are $\lesssim$ 0.01$\sim$0.02 $\text{yr}^{-1}$ and $\lesssim$ 0.04$\sim$0.08 $\text{yr}^{-1}$, respectively.
 \end{abstract}


\section{Introduction}

Gamma-ray bursts (GRBs), one of the most energetic explosions in the Universe,
are released during the rapid infall of material onto a newborn black hole or a neutron star,
originating from either a collapsar or a binary compact object merger \citep[see review by][]{Zhang18}.
The prompt emission of GRBs is thought to originate from the relativistic
jet ejected from the central compact remnant.
Owing to its ultra-relativistic motion, the prompt emission is highly beamed within a narrow angle
$\theta_{\text{jet}}+ 1/ \Gamma$, where $\theta_{\text{jet}}$ is the half opening angle of the jet,
$\Gamma$ is the initial Lorentz factor of the jet and $1/ \Gamma \ll \theta_{\text{jet}}$ \citep{Rhoads97,Huang00a,Huang00b,Huang00c,Granot02,Totani02,Zou07}.
If our line of sight is outside this angle, the GRB may be too faint to be detected.
When the jet slows down in the afterglow phase, the visible region gradually increases and could be observed over a wider angular range.
Such an afterglow not associated with any detectable GRBs is called an ``orphan afterglow'' \citep{Nakar02}.
While orphan afterglows are widely regarded as a beaming effect of GRB outflows, it should be noted that they could also come from failed GRBs, i.e. highly contaminated fireballs with a relatively smaller Lorentz factor \citep{Meszaros93,Meszaros94,Piran93,Huang02}.

The cosmological origin of GRBs indicates that a fraction of GRBs is likely to be gravitationally lensed by foreground objects during their propagation.
\cite{Paczynski86b} proposed that GRBs could be gravitationally lensed and multiple images of the same burst would be detected.
\cite{Mao92} studied the probability distributions of time delay in gravitational lensing by point masses and isolated galaxies,
and predicted that the probability of multiple GRB images due to galaxy lensing event is roughly between $0.05\%$ and $0.4\%$.
Assuming that 800 bursts per year are above the detection threshold of the Burst and Transient Source Experiment (BATSE),
\cite{Grossman94} predicted a rate of one lensing event caused by the known population of galaxies every $1.5$-$25$ years, and the corresponding median time delay between images is about $7$ days.
\cite{Nemiroff95} discussed the applications of gravitationally lensed GRBs in probing massive compact halo objects (MACHOs).
If GRBs detected by two detectors separated by approximately 1 AU have significantly different fluxes,
this would signal the presence of a MACHO $\leq 10^{-7}\ M_{\odot}$.
At the same time, several methods for distinguishing lensed GRBs have been proposed and discussed \citep{Wambsganss93,Nowak94}.
Based on observations from different dedicated detectors, a series of searches for lensed GRBs have been carried out in the gamma-ray band.
A search for lensed GRB pairs among 611 BATSE bursts produced no credible candidates \citep{Nemiroff94}.
\cite{Hanlon95} performed spectral and temporal analysis on GRB 930704 and GRB 940301,
and excluded the possibility that they were a lensed pair.
A search for lensed GRBs in the first-year observations of \emph{Fermi}/$\text{Gamma-Ray Burst Monitor}$
(GBM) yielded null results \citep{Veres09,Davidson11}.
Later, searches among GRBs observed by BATSE,
Konus-Wind, and GBM have been performed separately \citep{Li14,Hurley19,Ahlgren20},
no evidence for lensed signals with time delays within days was found.
Recently, \cite{Paynter21} claimed a statistically significant lensing event in the light curve of GRB 950830 based on Bayesian analysis,
from which they inferred that the lensing object was probably an intermediate-mass black hole.
\cite{Wang21} and \cite{Yang21} independently argued that GRB 200716C observed by GBM was likely a lensed GRB.
Four lensed GRB candidates are found in the search of the Fermi GBM data up to 2021 April \citep{Lin22}.
In addition, some efforts are made to investigate other lensing effects of GRB afterglows,
e.g., microlensing effects~\citep{Loeb98,Garnavich00}.

In general, there are some difficulties in identifying lensed candidates in the gamma-ray band based on the comparison of light curves.
The light curves of the majority of prompt GRB pulses are erratic and the radiation timescale is relatively short,
some real lensing events may be missed.
Moreover, the observation data of prompt emission are inevitably affected by the background noises and the positioning error of gamma-ray detectors is large,
making the shape of light curves be seriously altered.
Recently, \cite{Chen22} proposed that searching for lensed GRB events from the multi-band afterglow data could be promising,
in view of the much longer duration of afterglows and their simpler profiles of light curves.
However, only on-axis top-hat jet is discussed in this work.
Compared to ordinary afterglows detected in the follow-up observations after the GRB trigger,
the event rate of orphan afterglows should be higher.
Thus the orphan afterglow database may be a good reservoir for searching lensed events,
especially in the coming era of the time-domain survey.
On the other hand, the episode of short GRB 170817A in association with the gravitational wave event (GW170817, \citealt{Abbott17a})
shows that the GRB jet is structured \citep{Granot17,Lamb17,Meng19,Li19,Xiao17},
which is naturally expected from simulations \citep{Lazzati18,Bromberg18,Gottlieb18,Kathirgamaraju18,Geng19,Gottlieb21}.
Orphan afterglows produced by structured jets should be taken into account in relevant studies.

In this work, we discuss the gravitational lensing effect on multi-band orphan afterglows from structured jets.
A brief description of the GRB afterglow model is given in Sec. \ref{sec:afterglowmodel}.
In Sec. \ref{sec:lensmodel}, three gravitational lens models are described.
They include the point-mass (PM) model for objects that can be treated as a point-mass lens such as a Schwarzschild black hole,
the singular isothermal sphere (SIS) model for describing a galaxy with a specific mass distribution,
and the Chang-Refsdal (CR) model for a point mass lens perturbed by a galaxy with an external shear.
In Sec. \ref{sec:lightcurves}, the light curves of gravitationally lensed orphan afterglows are discussed.
The derivation of the redshift mass of the galactic gravitational lens from
lensed ophan afterglows using the autocorrelation function method is presented in Sec. \ref{sec:massestimation}.
The event rate for galaxy-lensed orphan afterglows is estimated in Sec. \ref{sec:eventrate},
while the detection rate with future survey telescopes is discussed in Sec. \ref{sec:detectrate}.
Finally, in Sec. \ref{sec:summary}, we summarize and discuss our results.
Throughout this paper a flat Lambda cold dark matter ($\Lambda$ CDM) cosmology model
with $H_{0}=71 \mathrm{~km} \mathrm{~s}^{-1} \mathrm{Mpc}^{-1}$, $\Omega_{\mathrm{m}}=0.27$, and $\Omega_{\Lambda}=0.73$ is adopted.

\vspace{-0.4cm}

\section{GRB afterglow model}
\label{sec:afterglowmodel}
Following the GW170817 from a binary neutron star merger,
a faint short burst GRB 170817A was detected \citep{Abbott17b,Goldstein17,Savchenko17}.
The time evolution of subsequent multi-wavelength afterglow emission is consistent with a relativistic off-axis structured jet \citep{Lazzati18,Mooley18b,Piro19},
and ruled out the uniform ``top-hat'' jet \citep{Alexander17,Haggard17,Margutti17,Murguia-Berthier17,Troja17}.
Other hydrodynamic \citep[e.g.,][]{Morson07,Mizuta09,Bromberg11,Lazzati15,Geng16b,Lopez16} and magnetohydrodynamic simulations \citep[e.g.,][]{Bromberg16}
also support that a structured jet is produced after its propagation inside a star envelope.
Thus, in this paper, we consider the jet structured to calculate afterglows.
For clarity, a power-law structured jet is assumed as (the case of Gaussian jet is given in Appendix \ref{sec:Gaussian}),
\begin{equation}
\begin{gathered}
\varepsilon(\theta)= \begin{cases}\varepsilon_{c}, & \theta<\theta_{c},\\
\varepsilon_{c}\left(\theta / \theta_{c}\right)^{-k_{e}}, & \theta_{c}<\theta<\theta_{m},\end{cases} \\
\Gamma(\theta)= \begin{cases}\Gamma_{c}, & \theta<\theta_{c}, \\
\Gamma_{c}\left(\theta / \theta_{c}\right)^{-k_{\Gamma}}+1, & \theta_{c}<\theta<\theta_{m},\end{cases}
\end{gathered}
\end{equation}
where $\theta_{c}, \varepsilon_{c}$, and $\Gamma_{c}$ are the half opening angle, the kinetic energy density, and the Lorentz factor of the inner core, respectively.
$\theta_{m}$ is the maximum of the half opening angle.
The indexes $k_{e}$ and $k_{\Gamma}$ describe the angular distribution of kinetic energy
density $\varepsilon(\theta)$ and Lorentz factor $\Gamma (\theta)$ within the jet cone.

Unlike the main mechanism of the prompt emission which is still under debate~\citep{Peer06,ZhangB11,Lundman13,Deng14,Uhm14,Geng18a,Gao21,Meng22},
afterglows are widely known as the synchrotron radiation of electrons accelerated by
the external shock generated from the interaction between the outflow and the circumburst medium~\citep{Peer12}.
During this process, a fraction ($\epsilon_{\text{e}}$) of the shock energy goes into electrons,
and a fraction ($\epsilon_{B}$) of the shock energy enters into the magnetic field.
The energy distribution of electrons accelerated by the external shock is usually assumed to be a power law, i.e.,
$dN_{\text{e}}/d\gamma_{\text{e}} \propto \gamma_{\text{e}}^{-p}$.
These electrons radiate synchrotron emission in the magnetic fields, and the broad-band afterglow is produced.
Without the consideration of lateral expansion of jets, the dynamical evolution of each jet segment tagged by different $\theta$ can be described by \citep{Huang99,Huang00c,Geng14,Geng16a},
\begin{equation}
\frac{d \Gamma}{d M_{\text{sw}}} = -\frac{\Gamma^{2}-1}{M_{\text{ej}} + \epsilon M_{\text{sw}}
+ 2 (1 - \epsilon) \Gamma M_{\text{sw}}},
\end{equation}
where $\epsilon$ is the radiation efficiency, $M_{\text{sw}}$ is the swept-up mass by the shock,
and $M_{\text{ej}}$ is the initial mass ejected from the central engine.
Integrating the emissions from all segments over the equal arrival time surfaces \citep{Waxman97,Sari98,Huang00b,Huang00c},
we could get the afterglow light curves of the structured jet.

\vspace{-0.4cm}

\label{sec:lensmodel}
\begin{figure}
  \centering\includegraphics[scale=.7]{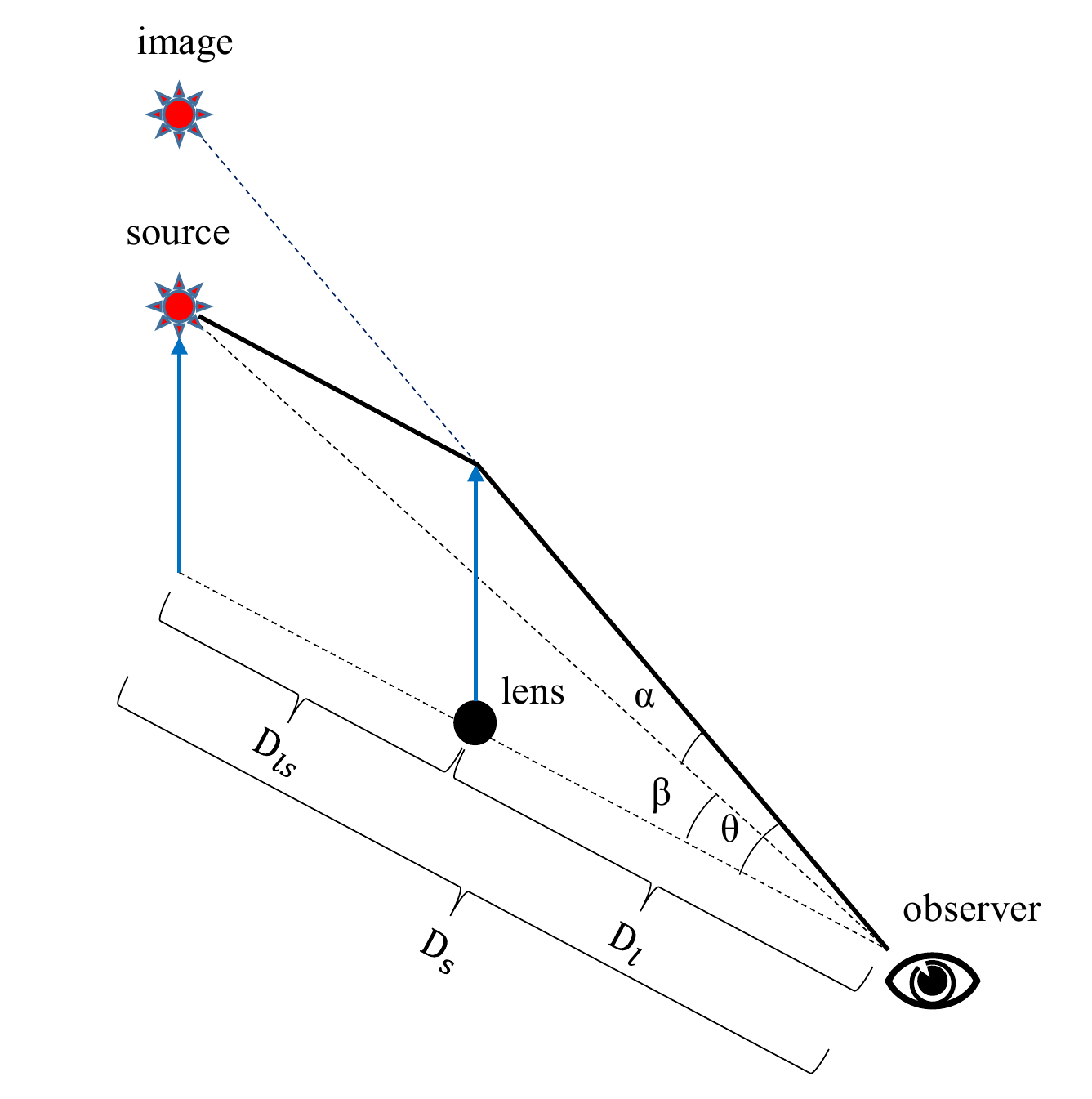}
    \caption{The geometry of the PM model and the SIS model.}
    \label{PMmodel}
\end{figure}

\section{Gravitational lens model}
\label{sec:lensmodel}
Here, we briefly introduce three prevailing lens models, i.e., the PM model, the SIS model, and the CR model.
The gravitational effect of a compact star (e.g., a black hole or a neutron star) can be described by the PM model \citep{Schneider92,Virbhadra98,Virbhadra02}.
The Einstein angle corresponding to a point mass $M_{l}$, is defined as
\begin{equation}
\theta_{E}=\sqrt{\frac{4 G M_{l}}{c^{2}} \frac{D_{l s}}{D_{l} D_{s}}},
\end{equation}
where $D_{l}, D_{s}$, and $D_{l s}$ are the angular diameter distances from the lens to the observer, from the source to the observer, and from the source to the lens respectively (see Fig. \ref{PMmodel}).
The lens potential of a point-mass lens model is
\begin{equation}
\psi(\theta)= \ln (\theta),
\end{equation}
and the lens equation scaled by $\theta_{E}$ is
\begin{equation}
y=x-\frac{1}{x},
\label{eq:lenseq1}
\end{equation}
where $y=\beta / \theta_{E}$, $x=\theta / \theta_{E}$.
$\beta$ and $\theta$ are angular positions of the source and image, respectively (see Fig. \ref{PMmodel}).
Generally, the lens equation has two solutions, i.e., $x_{\pm}=\frac{1}{2}\left(y \pm \sqrt{y^{2}+4}\right)$,
where ``+'' and ``-'' denote the parity of the image.
The time delay between these two images is
\begin{equation}
\Delta t=\frac{4 G M_{l}}{c^{3}}\left(1+z_{l}\right)\left[\frac{y \sqrt{y^{2}+4}}{2}+\ln \left(\frac{\sqrt{y^{2}+4}+y}{\sqrt{y^{2}+4}-y}\right)\right],
\end{equation}
where $z_{l}$ is the redshift of the lens, and the positive parity image is always the leading image.
Their magnifications are
\begin{equation}
\mu_{\pm}=\pm \frac{1}{4}\left[\frac{y}{\sqrt{y^{2}+4}}+\frac{\sqrt{y^{2}+4}}{y} \pm 2\right].
\end{equation}

\begin{figure}
	\centering\includegraphics[scale=.5]{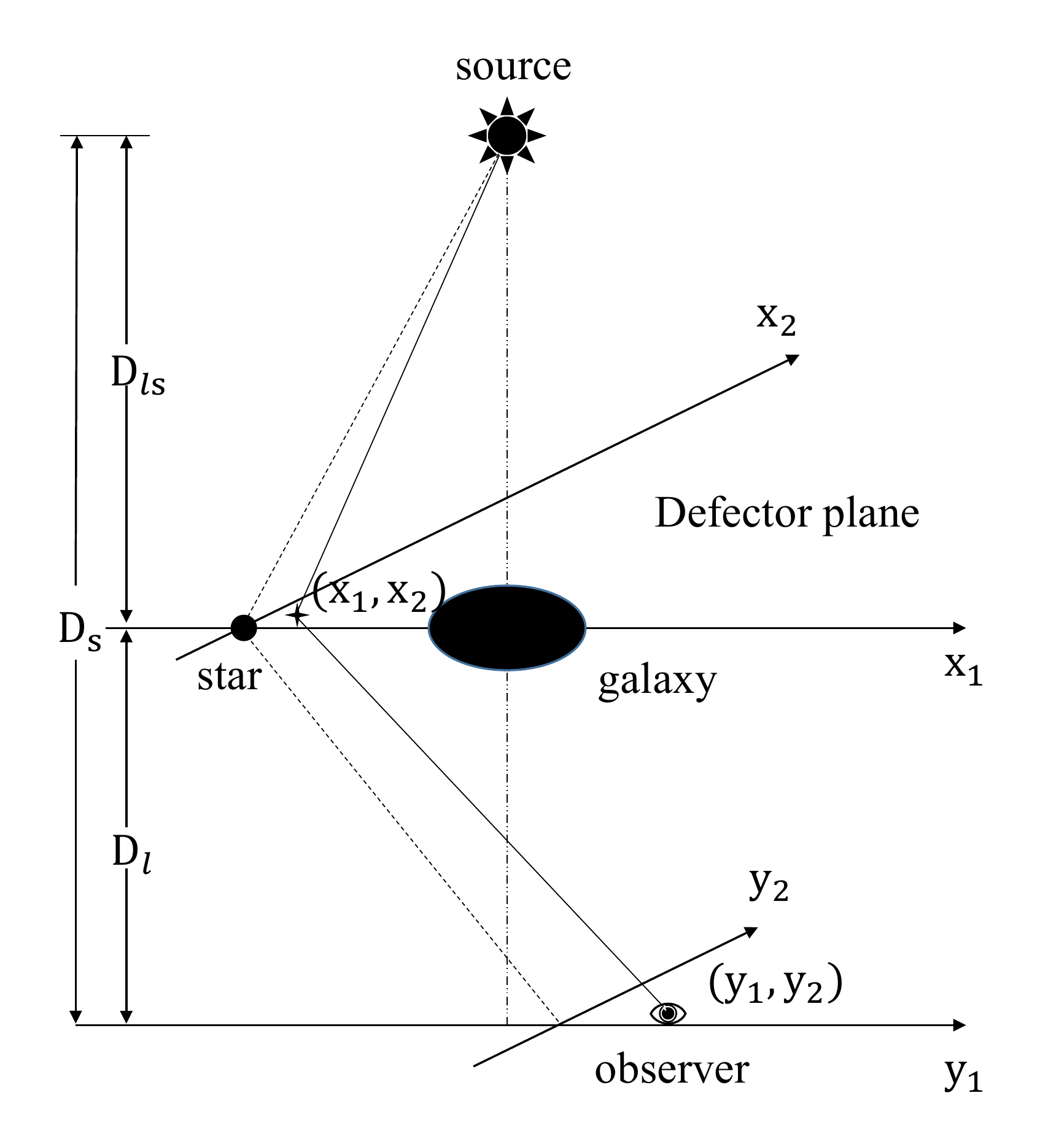}
    \caption{The geometry of the Chang-Refsdal model.
    A similar figure and a detailed description could be found in \citet{Chang79}.}
    \label{CRmodel}
\end{figure}

The SIS model is often used to describe a galaxy acting as the gravitational lens \citep{Schneider92} (see Fig. \ref{PMmodel}).
Its Einstein angle is defined as,
\begin{equation}
\theta_{E}=\sqrt{\frac{4 G M(\theta_{E})}{c^{2}} \frac{D_{l s}}{D_{l} D_{s}}}=4 \pi \frac{\sigma_{v}^{2}}{c^{2}} \frac{D_{l s}}{D_{s}},
\end{equation}
where $M(\theta_{E})$ is the mass within the Einstein radius, and $\sigma_{v}$ is the velocity dispersion of the galaxy.
The corresponding scaled lens equation is,
\begin{equation}
y=x-\frac{x}{|x|}.
\end{equation}
When $y<1$, there are two solutions: $x_{\pm}=y\pm 1$.
However, when $y>1$, the gravitational effect is weak and the lens equation has only one solution: $x=y+1$,
hard to justify the existence of the lens.
Here, we focus on the case of $y<1$.
In this case, the time delay between different images reads
\begin{equation}
\Delta t=\frac{32 \pi^{2}}{c}\left(\frac{\sigma_{v}}{c}\right)^{4} \frac{D_{l} D_{l s}}{D_{s}}\left(1+z_{l}\right) y,
\label{eq:SISTD}
\end{equation}
and the magnifications of the images are
\begin{equation}
\mu_{\pm}=\mid1 \pm \frac{1}{y}\mid.
\label{eq:SISmagnification}
\end{equation}

\begin{figure*}
	\vskip-0.1in
    \includegraphics[angle=0,scale=0.7]{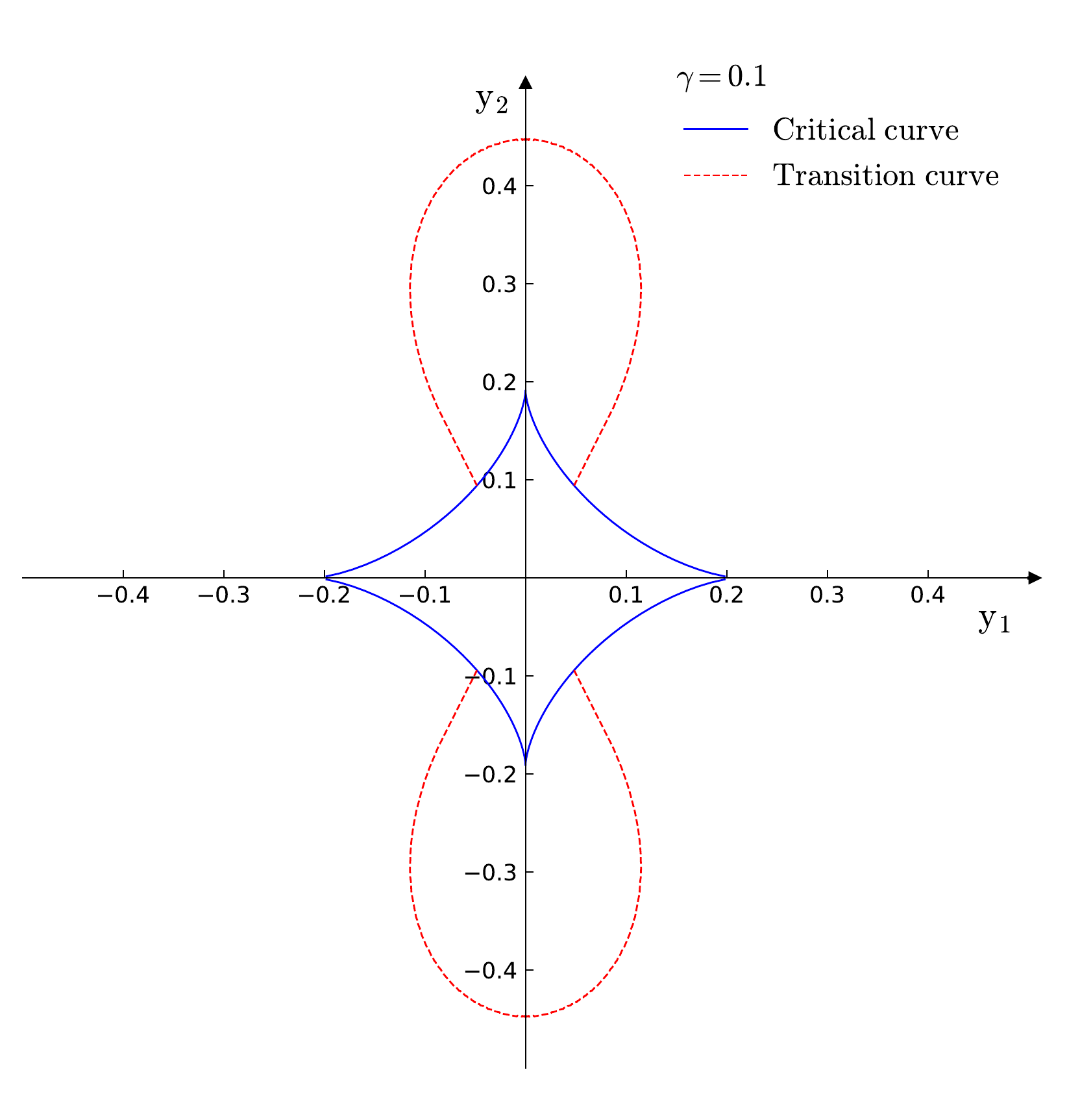}
    \caption{
    Critical curves (the blue solid line) and transition curves (the red dotted line) in the observer's plane ($y_{1}$, $y_{2}$) for the case $\gamma=0.1$.
    The observer can see four images inside the range of the critical curve and can see two images outside this range.
    In particular, when the observer is right located on the coordinate axis ($y_{1}$-axis or $y_{2}$-axis) within the range of the critical curve,
    two of four images will reach the observer at the same time.
    On the other hand, a brighter trailing image will be observed inside the transition curve range,
    while a brighter leading image will be detected outside the transition curve.
    }
    \label{CriticalCurve}
\end{figure*}

The CR model (see Fig. \ref{CRmodel}) shows a point mass lens perturbed by a galaxy with an external shear \citep{Chang79,Chang84,Schneider92,Chen21a},
and its effective lensing potential is
\begin{equation}
\psi(\theta)= \ln (\theta)-\frac{\gamma}{2}\left(\theta_{1}^{2}-\theta_{2}^{2}\right),
\end{equation}
where $\gamma$ is the external shear strength.
The lens equation in dimensionless form is:
\begin{equation}
\begin{aligned}
&y_{1}=(1+\gamma) x_{1}-\frac{x_{1}}{x_{1}^{2}+x_{2}^{2}}, \\
&y_{2}=(1-\gamma) x_{2}-\frac{x_{2}}{x_{1}^{2}+x_{2}^{2}}.
\end{aligned}
\end{equation}
The lens equation has four possible solutions at most and needs to be solved numerically.
When the observer stands in different positions ($y_{1}$, $y_{2}$), different numbers of images will be seen (see Fig. \ref{CriticalCurve}).
The magnification of each image (to the unlensed one) writes as
\begin{equation}
\mu=(1-\gamma^{2}-\frac{1}{|x|^{4}}-2 \gamma \frac{x_{1}^{2}-x_{2}^{2}}{|x|^{4}})^{-1}.
\end{equation}
The time delay between each two images is determined by
\begin{equation}
\Delta t=t(\vec{x_{s}})-t(\vec{x_{f}}),
\end{equation}
and
\begin{equation}
\begin{aligned}
&t(\vec{x})=\frac{4 G M_{l}}{c^{3}} (1+z_{l}) [\frac{1}{2} (\vec{x}-\vec{y})^{2}-\ln \lvert \vec{x} \rvert+ \frac{\gamma}{2} (\vec{x}^{2}-\vec{y}^{2})],
\end{aligned}
\end{equation}
where $\vec{x_{f}}$ and $\vec{x_{s}}$ are positions of the leading image and the trailing image, respectively.

\begin{figure*}
\vskip-0.1in
\includegraphics[angle=0,scale=0.55]{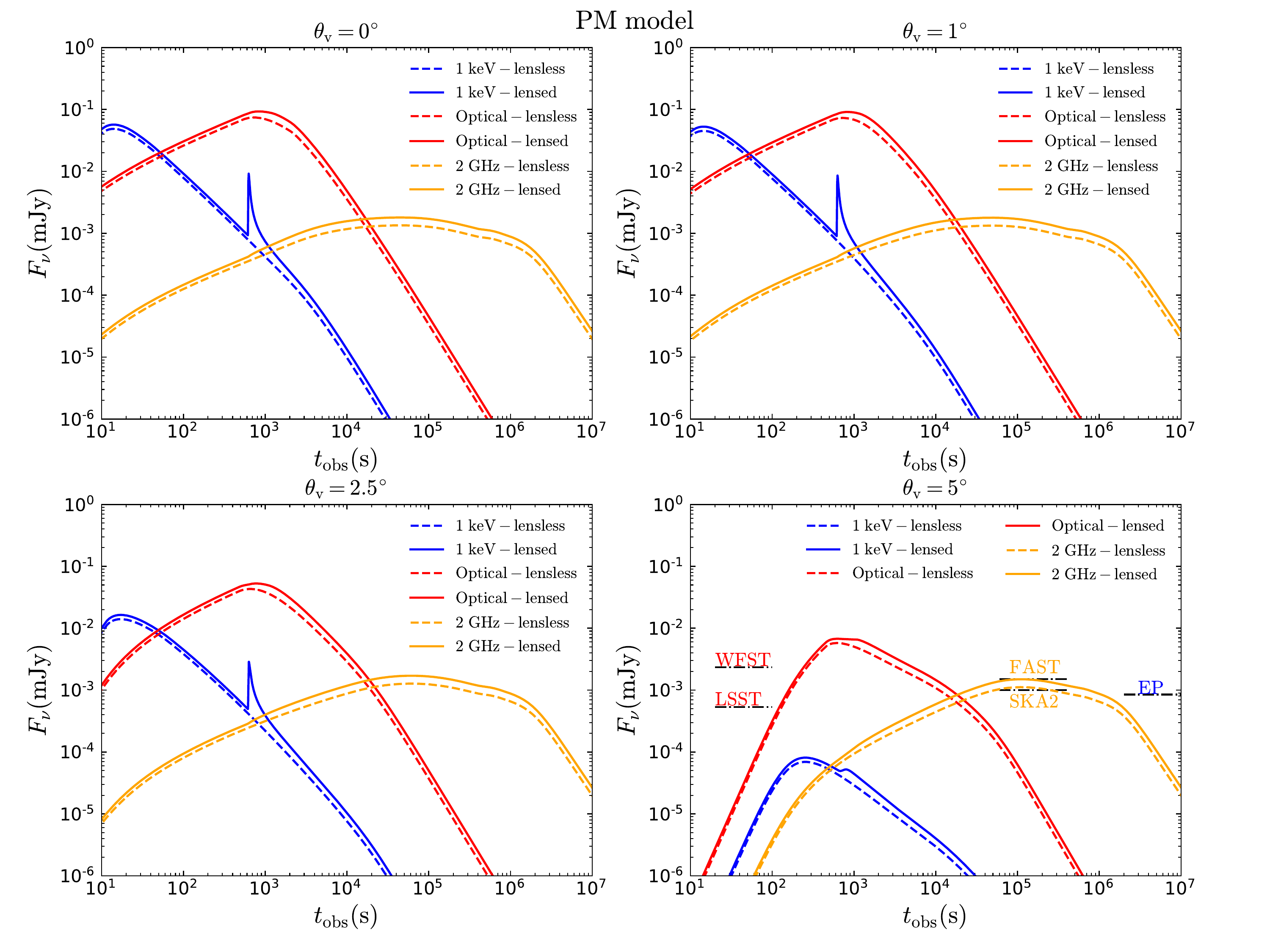} 
\caption{Lensed afterglow light curves for the PM model at different viewing angles $\theta_{\text{v}}$.
    Three bands are shown, including the X-ray band (blue), the optical band (red), and the radio band (orange).
    In each panel, the dashed lines are light curves without gravitational lensing, and solid lines represent lensed light curves.
    The horizontal dash-dot lines represent the sensitivities of the Five-hundred-meter Aperture Spherical radio Telescope (FAST), Square Kilometre Array Phase 2 (SKA2),
   Wide Field Survey Telescope (WFST), Vera Rubin Observatory Legacy Survey of Space and Time (LSST), and Einstein Probe (EP).
   }
    \label{PM}
\end{figure*}

\begin{figure*}
\vskip-0.1in
\includegraphics[angle=0,scale=0.55]{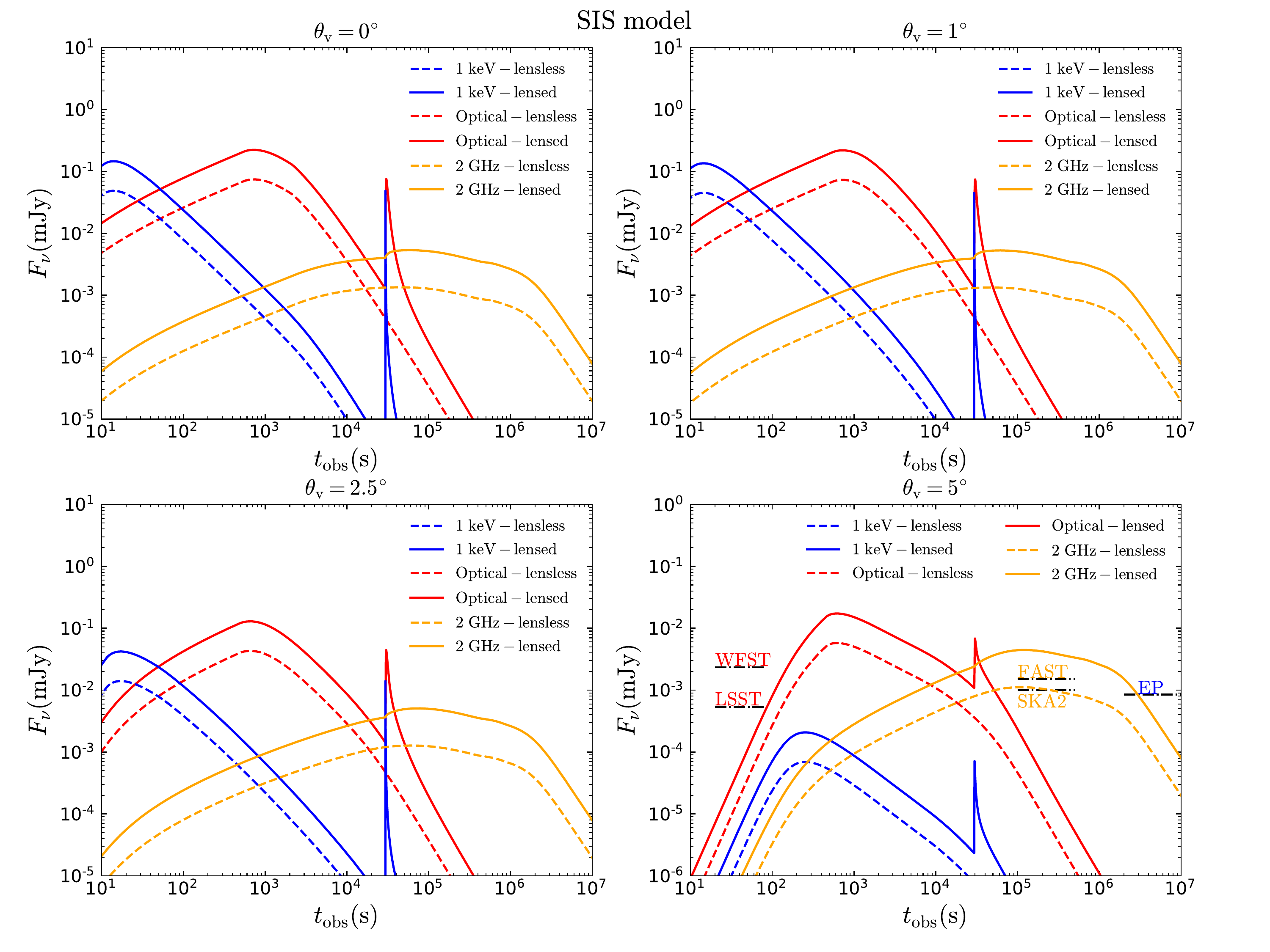}
    \caption{Lensed afterglow light curves for the SIS model.
    Line styles are the same as Fig. \ref{PM}.
    }
    \label{SIS}
\end{figure*}

\begin{figure*}
\vskip-0.1in
\includegraphics[angle=0,scale=0.55]{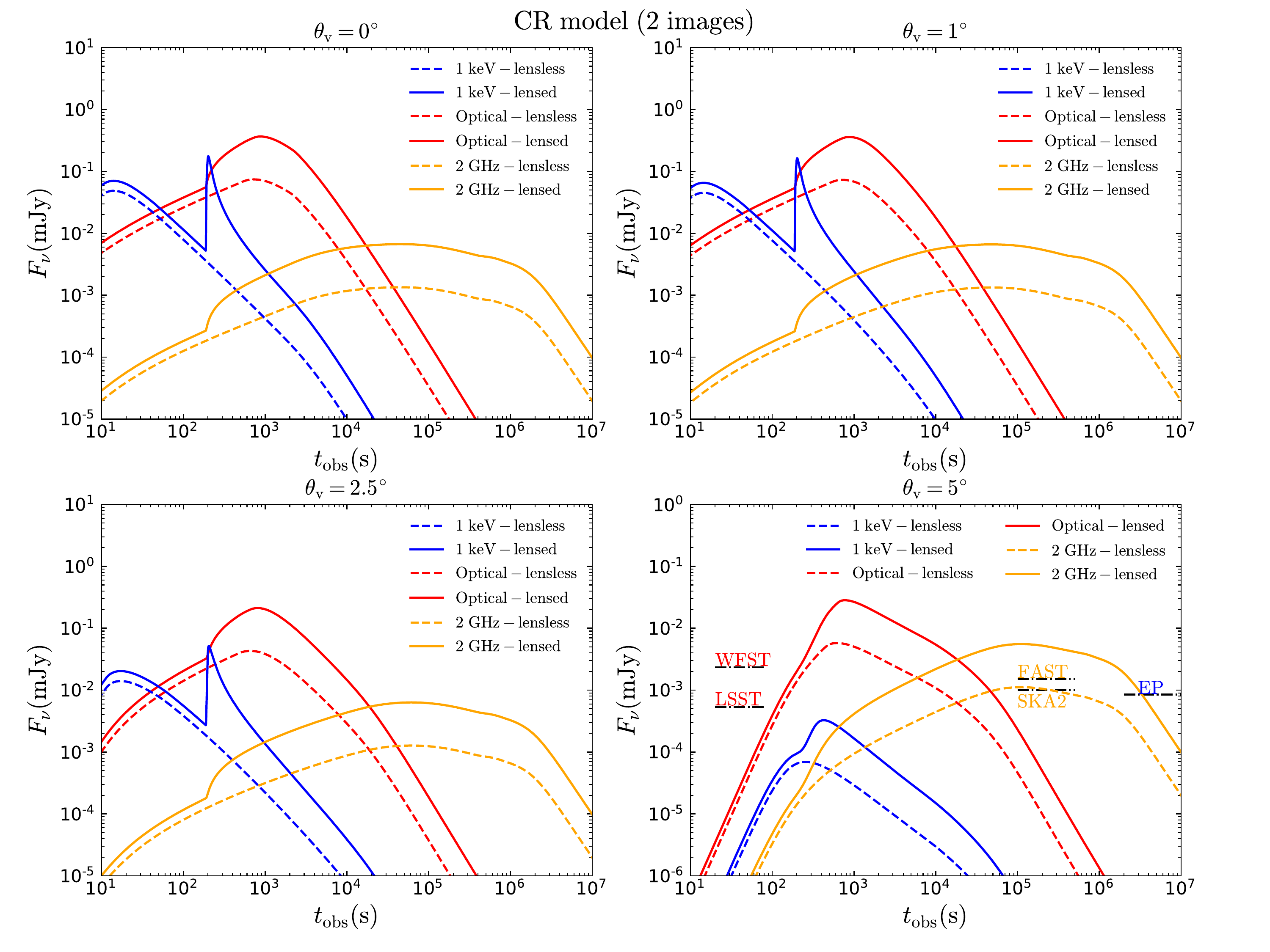}
    \caption{Lensed afterglow light curves for the CR model with two images.
    Line styles are the same as Fig. \ref{PM}.}
    \label{CR1}
\end{figure*}

\begin{figure*}
\vskip-0.1in
\includegraphics[angle=0,scale=0.55]{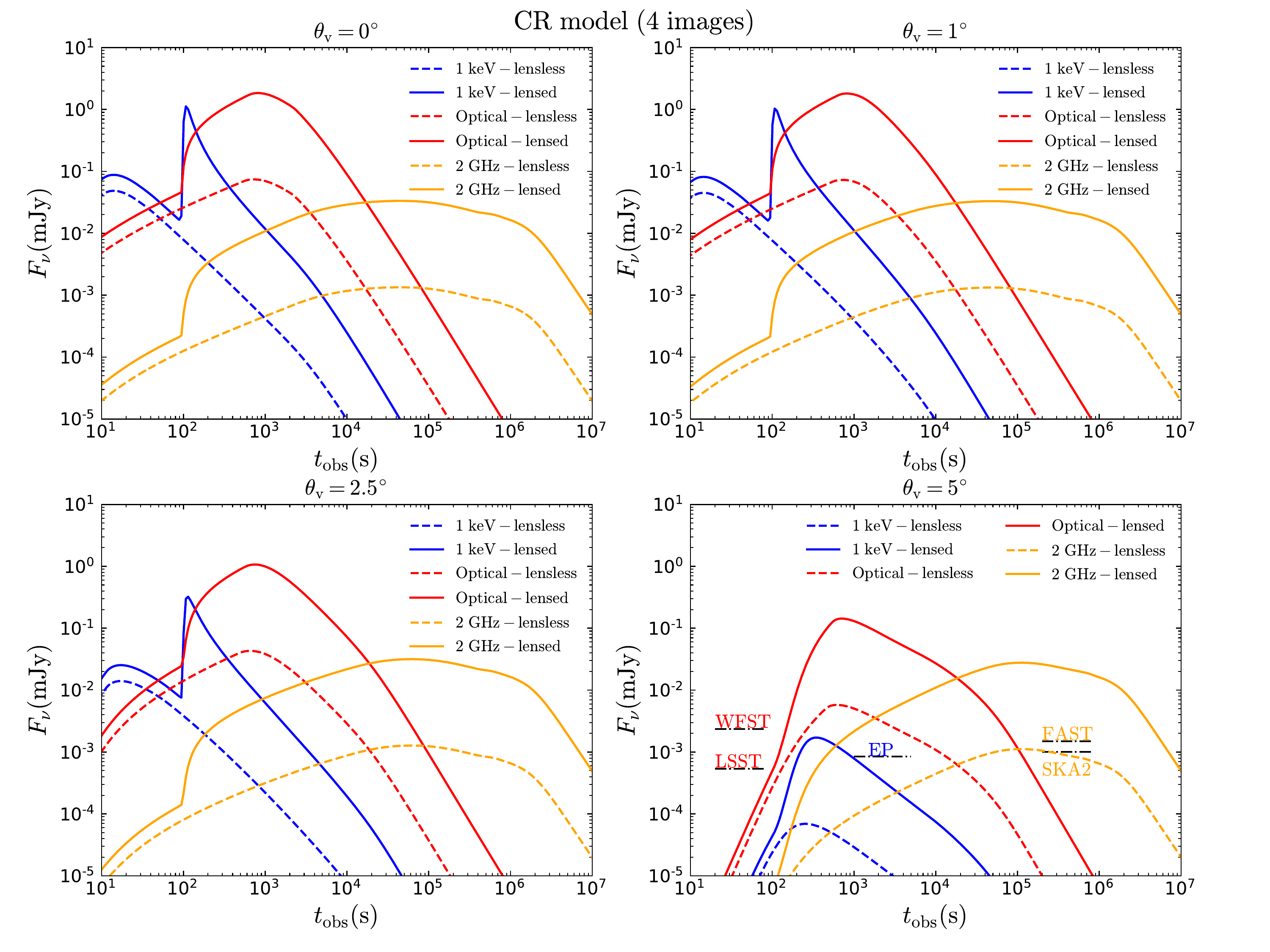}
    \caption{Lensed afterglow light curves for the CR model with three images.
    Line styles are the same as Fig. \ref{PM}.}
    \label{CR2}
\end{figure*}

\begin{figure*}
\vskip-0.1in
\includegraphics[angle=0,scale=0.55]{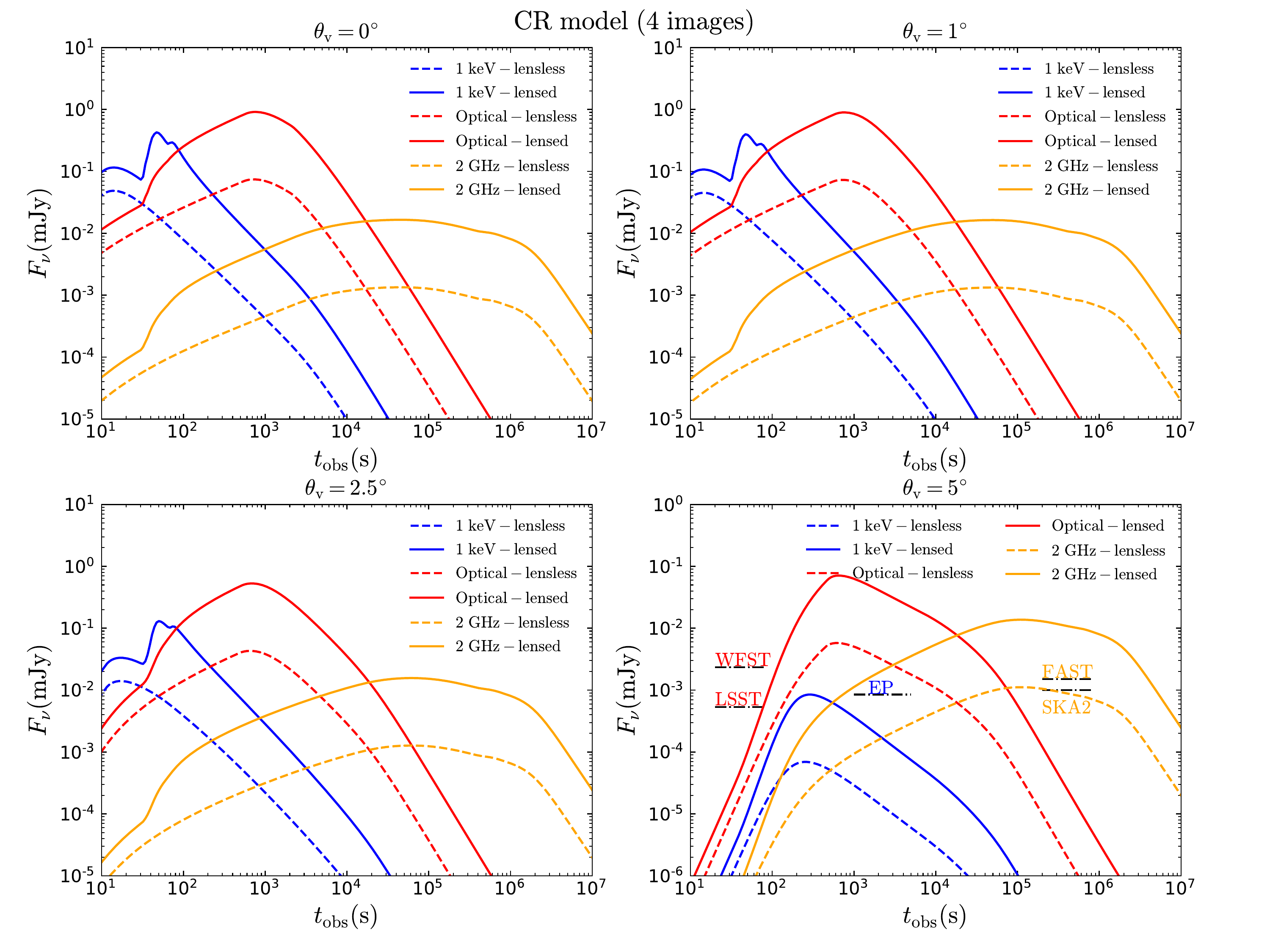}
    \caption{Lensed afterglow light curves for the CR model with four images.
    Line styles are the same as Fig. \ref{PM}.}
    \label{CR3}
\end{figure*}

\section{Lensed afterglow light curves}
\label{sec:lightcurves}

Using the afterglow calculation scheme mentioned above,
we first calculate the light curves in three different bands, including the X-ray band ($1$ keV), the optical band ($1.65$ eV), and the radio band ($2$ GHz).
Then, the corresponding lensed light curves are derived from different lens models.
In our calculations, the redshifts of GRB and gravitational lens are fixed as $z_{s}=1.0$ and $z_{l}=0.5$, respectively.
The half-opening angle of the inner core is adopted as $\theta_{c}=2^{\circ}$,
with an isotropic equivalent kinetic energy of $E=10^{52}$ erg,
and the maximum of the half opening angle is taken as $\theta_{m}=12^{\circ}$.
The index values for the power-law structure of the jets are set as $k_{\text{e}} = k_{\Gamma}=2.0$.
The viewing angle between the jet axis and the line of sight is taken as $\theta_{\text{v}}=0, 1^{\circ}, 2.5^{\circ}, 5^{\circ}$.
As $\theta_{\text{v}}$ becomes larger than $\theta_{c}$, the corresponding GRB prompt emission gets too weak to be detected,
making the afterglow to be an orphan afterglow.
Meanwhile, we adopt the initial Lorentz factor of the inner core as $\Gamma_{c}=300$.
Other microphysical parameters characterizing the energy fraction of electrons and magnetic field are set as $\epsilon_{\text{e}}=0.1$ and $\epsilon_{B}=0.01$.
The number density of the ambient medium is typically taken as $n=1.0\ \text{cm}^{-3}$.
The spectrum index of accelerated electrons is adopted as $p=2.5$.

For calculations of the PM model, the lens parameters are set as: $M_{l}=10^{7}\ M_{\odot}$, $\beta=1.0 \theta_{\text{E}}$,
from which the time delay of two images is expected to be about $10^{3}$~s,
and the flux ratio, defined as the ratio between magnifications of the leading image and the trailing image, is roughly $7$.
Figure \ref{PM} shows the lensed multi-band afterglows made of the superposition from two successively arriving images.
It could be seen that the peak of the later image will arise in the X-ray band
for $\theta_{\text{v}} \le \theta_{c}$.
In cases of $\theta_{\text{v}} > 2 \theta_{c}$, with a relatively late peak time for the first image and a shallow decay behavior after the peak,
determined solely by the jet dynamic and the viewing angle, the second peak will be smoothed as a shallow plateau.
Therefore, from on-axis afterglows to orphan afterglows,
the second image turns from a steep bump to a shallow plateau in the lensed X-ray light curves as $\theta_{\text{v}}$ increases.
In the optical and radio band, since the peak time of the first image is always larger than that of the delay time between two images,
it is hard to distinguish the flux contribution from each other in the rising phase of the light curve.
Hence no lensed signal will be clearly observed in these two bands.

In our calculations of the SIS model, the parameters for the lens model are: $M_{l}=10^{9}\ M_{\odot}$, $\beta=0.5\theta_{\text{E}}$.
Owing to a larger time delay induced by a larger lens mass, the second image will superpose on the late decay stage of the first image,
resulting in a spike-like bump in both the X-ray and the optical band (see Fig. \ref{SIS}).
This significant bump still exists for large $\theta_{\text{v}}$.
Similar to those in the PM model, no lensed signal could be clearly identified in the radio band.

For the CR model, the lens mass is set as $M_{l}=10^{7}\ M_{\odot}$, and the external shear is fixed as $\gamma=0.1$.
According to the observer's position with respect to the critical/transition curve, three cases are considered.
In Fig. \ref{CR1}, the position of observer in $y_{1}$-$y_{2}$ plane is [0, $0.3\theta_{\text{E}}$],
which is inside the transition curve (see Fig. \ref{CriticalCurve}).
Notably, the trailing image is brighter than the leading image, and the flux ratio is less than $1.0$.
It implies that another higher peak may arise after the first one,
which is exhibited by the X-ray light curves in Fig. \ref{CR1}.
Such a lensed signature could also appear when $\theta_{\text{v}}$ equals $5^{\circ}$.
Compared to the PM model, the CR model presets a more significant
lensed signature owing to the external shear from the galaxy.
In Fig. \ref{CR2}, it is assumed that the observer is at [0, $0.15\theta_{\text{E}}$], which is inside the critical curve range.
However, the second, third, and fourth images reach the observer almost simultaneously.
As a result, the second peak is strengthened and more obvious gravitational lensing signals emerge.
However, the coincidence of the coordinate axis and the observer position is very rare.
For an observer who stands inside the critical curve range with a position ($0.05\theta_{\text{E}}$, $0.05\theta_{\text{E}}$),
four images with different time delays will be received as shown in Fig. \ref{CR3}.
Since the time separation between the second and the third images is very short,
only three peaks appear in the X-ray light curves.
For the case of large $\theta_{\text{v}}$, the temporal index in the rising phase
evolves from the dynamically determined value to
a steeper value caused by the flux contribution of later images.

According to the results for lensed light curves under the PM, the SIS, and the CR model,
we argue that the lensed optical/X-ray orphan afterglows could be in principle diagnosed through
their temporal characteristics.
The time delay produced by a galaxy gravitational lens (SIS model) is large enough to generate a clear lensed signal in orphan afterglow light curves.
In the next section,
the redshift mass $(1+z_{l}) M_{l}$ for the galaxy lens will be estimated from such observational data.

\vspace{-0.4cm}

\section{Lens mass estimation}
\label{sec:massestimation}

Based on the time delay and the flux ratio, one can roughly estimate the redshift mass of the galaxy lens in the rest frame \citep{Mao92,Paynter21,Kalantari21,Chen21a}.
Combining Eq. \ref{eq:SISTD} and Eq. \ref{eq:SISmagnification}, we can get a formula for lens mass estimation of the SIS model,
\begin{equation}
(1+z_{l})M_{l}=\frac{c^{3}}{8G}\frac{R+1}{R-1} \Delta t,
\end{equation}
where $R=\mu_{+}/\mu_{-}$.
As is shown above, the mass of the lens can be determined from the observables ($\Delta t$, $R$, and $z_{l}$).
Generally, the redshift of the lens may be uncertain for some reasons,
so only the redshift mass of the lens could be derived.
The uncertainty of the redshift mass is contributed by two quantities, i.e., $\Delta t$ and $R$.
For the time delay $\Delta t$, it has $\frac{\delta M}{M}=\delta \Delta t / \Delta t$.
This uncertainty could be ignored when the time resolution of the detector is
much less than the time delay between the two images.
For the flux ratio $R$, one can derive $\frac{\delta M}{M}=\frac{2R}{R^{2}-1}\frac{\delta R}{R}$.
The first term on the right hand is about $1.3$ for $R=2.0$ and quickly drops to below 1 when $R > 2.4$.
Therefore the relative uncertainty in redshift mass is comparable with $\delta R / R$ and decreases with increasing $R$.
In short, the measured flux ratio is mainly responsible for errors in the redshift mass estimation.

\begin{figure*}
\vskip-0.1in
\includegraphics[angle=0,scale=.8]{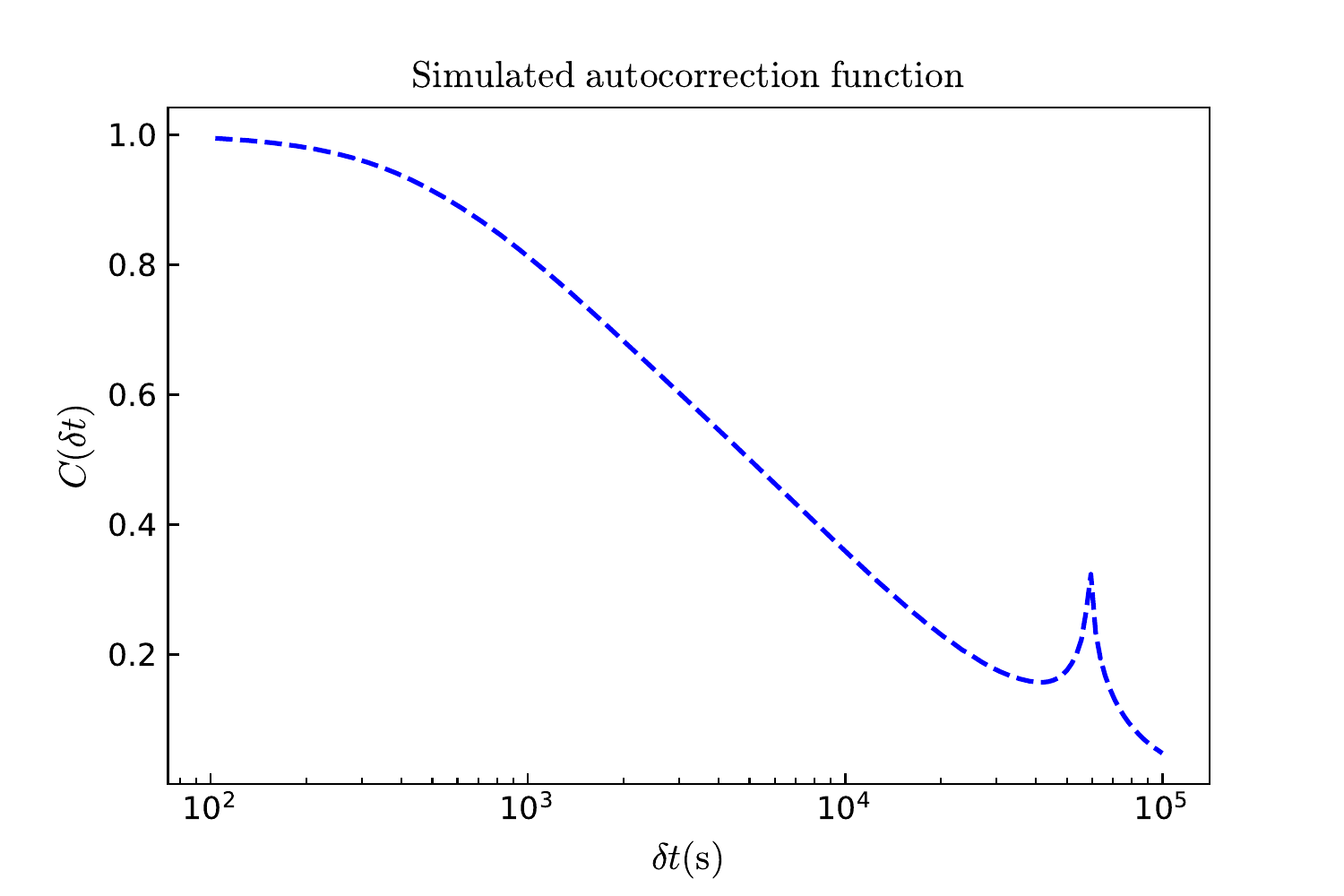}
    \caption{The simulated results of the autocorrelation function for lensed orphan afterglows of the SIS model.}
    \label{auto}
\end{figure*}

However, the lensed orphan afterglow light curves are the results of the superposition of two images, and the flux ratio could not be directly derived.
Here, we adopt the autocorrelation function to obtain the time delay and flux ratio.
The autocorrelation function is defined by \citep{Hirose06,Ji18},
\begin{equation}
C(\delta t) \equiv \frac{\int d t I(t) I(t-\delta t)}{\sqrt{\int d t I^{2}(t)} \sqrt{\int d t I^{2}(t-\delta t)}},
\end{equation}
where $I(t)$ is the light curve and note that $C(0)=1$.
For generic uncorrelated functions, $C(\delta t \neq 0)<1$, and thus $\delta t=0$ serves as the only major spike in the function $C(\delta t)$.
Let $\hat{I}(t)$ be the intrinsic light curve of the orphan afterglow and $\hat{C}(t)$ its autocorrelation.
It is easy to get,
\begin{equation}
I(t) \propto \frac{R}{R+1} \hat{I}(t)+\frac{1}{R+1} \hat{I}(t-\Delta t),
\end{equation}
and
\begin{equation}
\begin{aligned}
C(\delta t)=\frac{(R^{2}+1) \hat{C}(\delta t)+R[ \hat{C}(\delta t+\Delta t)+ \hat{C}(\delta t-\Delta t)]}{(R^{2}+1)+2 R  \hat{C}(\Delta t)}.
\end{aligned}
\end{equation}
If we assume $\hat{C}(\Delta t)=0$ and $\hat{C}(2\Delta t)=0$ as an approximation,
the lensed signal will show spikes at $\delta t=-\Delta t, 0, \Delta t$ with an amplitude ratio $R/(R^{2}+1)$:$1$:$R/(R^{2}+1)$.
Moreover, the flux ratio can be derived from this amplitude ratio of spikes in the autocorrelated function.
Here, taking the lensed orphan afterglow light curves of the SIS model ($M_{l}=2\times10^{9}\ M_{\odot}$, $\theta_{\text{v}} = 5^{\circ}$) as an example,
we perform autocorrelation function simulations and the results are shown in Fig. \ref{auto}.
In practice, the observed light curves are composed of discrete data points, so the observation time resolution will affect the estimation of the redshift mass.
To achieve good accuracy of the estimation, some limitations on the time resolution of observation are required.
First, the time resolution is required to be smaller than $10^{4}$ s which is comparable to the time delay between the two images caused
by a galaxy lens ($\geq 10^{9}\ M_{\odot}$). 
Moreover, the time resolution has to be less than the peak time ($\sim10^3$ s) of the intrinsic light curves of orphan afterglows, for observing the first bump
caused by the leading image in the overlapped light curve.
As a result, it is found that the relative error between the real redshift mass and the estimated value will be less than 10\% if the time resolution is less than 330 s.

\section{Event rates for galaxy-lensed orphan afterglows}
\label{sec:eventrate}
We now estimated the event rate for galaxy-lensed orphan afterglows with the SIS model.
A source with an angular impact parameter $\beta$ has an effective lensing cross-section of \citep{Paynter21},
\begin{equation}
\begin{aligned}
\int d\sigma &=\int_{\beta_{\min }}^{\beta_{\max }} 2 \pi \beta d \beta \\
&=\pi \theta_{E,SIS}^{2} \int_{y_{\min }}^{y_{\max }} 2 y d y \\
&=\frac{4 \pi G M_{l,SIS}}{c^{2}} \frac{D_{ls}}{D_{l}D_{s}} \int_{y_{\min }}^{y_{\max }} 2 y d y ,
\end{aligned}
\end{equation}
where $y_{\min }$ is determined by Eq. \ref{eq:SISTD} in which the time delay $\Delta t$ is $1000$ s, and $y_{\max }=1.0$.
Notably, it is assumed that the minimum mass of the lensing galaxy is $10^{9} M_{\odot}$.
The resulting cross section is \citep{Paynter21},
\begin{equation}
\sigma(z_{s},z_{l},M_{l,SIS})=\frac{4 \pi G M_{l,SIS}}{c^{2}} \frac{D_{ls}}{D_{l}D_{s}}\left(y_{\max }^{2}-y_{\min }^{2}\right) \Theta\left(y_{\max }-y_{\min }\right),
\end{equation}
where $ \Theta$ is the Heaviside step function.
The optical depth denotes the probability for a GRB to be lensed, and is written as \citep{Vietri83,Turner84,Paczynski86a,Nemiroff89,Munoz16,Paynter21},
\begin{equation}
\tau(z_{s})=\frac{1}{d \Omega} \int_{0}^{z_{s}} d V\left(z_{l}\right) \int d M_{l,SIS} n\left(z_{l},M_{l,SIS}\right) \sigma(z_{s},z_{l},M_{l,SIS}) ,
\end{equation}
where $n\left(z_{l},M_{l,SIS}\right)$ is the comoving density of lenses, and $d V(z_{l})$ is the comoving volume element which is defined by
\begin{equation}
\begin{aligned}
d V(z) &=\chi^{2}(z) \frac{d \chi(z)}{d z} d z d \Omega \\
&=\chi^{2}(z) \frac{c}{H_{0}} \frac{d z d \Omega}{\sqrt{\Omega_{\Lambda}+\Omega_{\text{m}}(1+z)^{3}}},
\end{aligned}
\end{equation}
where $\chi(z)$ is the comoving distance corresponding to $z$.

The event rate is calculated by
\begin{equation}
N= \int_{0}^{z_{s,max}} \frac{\Delta \Omega-\Delta \Omega_{0}}{\Delta \Omega_{0}} N_{0}f\left(z_{s}\right) \tau(z_{s}) d z_{s},
\end{equation}
where $N_{0}$ is the average number of GRBs detected by \emph{Swift}/BAT per year, $f\left(z_{s}\right)$ denotes the observed GRB redshift distribution in unit of $\text{sr}^{-1}$, $\Delta \Omega$ is the corresponding solid angle to the maximum of viewing angle $\theta_{\text{v}}$, and $\Delta \Omega_{0}$ is the solid angle corresponding to the jet inner core $\theta_{c}$, and $(\Delta \Omega-\Delta \Omega_{0}) /\Delta \Omega_{0}$ is the ratio of numbers between orphan afterglows and ordinary afterglows which are associated with GRBs.

Based on \cite{Mortlock15}, we derive the comoving density of galaxies whose mass is greater than $10^{9}\ M_{\odot}$ from $z_{l}=0.3$ to $z_{l}=2$.
In the range of $0.3<z<3$, the corresponding total galaxy stellar mass function with a Schechter function form is,
\begin{equation}
\begin{gathered}
\phi(M)= \begin{cases}\ln (10) \cdot \exp \left[-10^{\left(M-M^{*}\right)}\right] \cdot 10^{\left(M-M^{*}\right)}\\
\cdot[\phi_{1}^{*} \cdot 10^{\left(M-M^{*}\right) \alpha_{1}}
+\phi_{2}^{*} \cdot 10^{\left(M-M^{*}\right) \alpha_{2}}],
& 0.3<z<1,\\
\phi^{*} \cdot \ln (10) \cdot\left[10^{\left(M-M^{*}\right)}\right]^{(1+\alpha)} \cdot \exp \left[-10^{\left(M-M^{*}\right)}\right],
& 1<z<3,\end{cases} \\
\label{eq:SMF}
\end{gathered}
\end{equation}
where $\phi^{*}$ is the normalization of the Schechter function, $M^{*} $ is the
turnover mass in units of index, $\alpha$ is the slope of the low-mass end
of the Schechter function and $M$ is the stellar mass in units of index.
The parameters, including $\phi^{*}$, $M^{*} $, and $\alpha$, vary with the range of the redshift $z_{l}$,
and could be found in Table 2 and 3 of \cite{Mortlock15}.

Using the luminosity evolution model and the corresponding parameters in \cite{Lan21}, we get the redshift distribution of long GRBs detected by \emph{Swift}/BAT from $z_{s}=0$ to $z_{s}=10$ with,
\begin{equation}
\begin{split}
&f(z_{s})=\frac{c}{H_{0}} \frac{\chi^{2}(z_{s})}{\sqrt{\Omega_{\Lambda}+\Omega_{\text{m}}(1+z_{s})^{3}}}\cdot\\ &\int_{\max \left[L_{\min }, L_{\lim }(z_{s})\right]}^{L_{\max }} \Theta(P(L, z_{s})) \frac{\psi(z_{s})}{1+z_{s}} \phi(L, z_{s}) \mathrm{d} L,
\end{split}
\end{equation}
where $\Theta(P)$ is the detection efficiency,
$P$ is the peak flux of the burst,
$\psi(z)$ is the comoving event rate of GRBs in units of $\text{Mpc}^{-3} \text{yr}^{-1}$,
$L$ is the luminosity of GRBs,
$\phi(L, z)$ is the normalized GRB luminosity function,
and more details could be found in \cite{Lan21}.

For an assumed maximum viewing angle of $\theta_{\text{v}}=5^{\circ}$,
the final result for the lensed event rate is $\lesssim 0.15\ \text{yr}^{-1} \text{sr}^{-1}$,
and the lensed event rate for the whole sky is $\lesssim 1.8\ \text{yr}^{-1}$.

\vspace{-0.4cm}

\section{Detection rates of galaxy-lensed orphan afterglows}
\label{sec:detectrate}

In the lower right panel of Fig. \ref{SIS}, the sensitivity of several detectors, including the Five-hundred-meter Aperture Spherical radio Telescope (FAST), Square Kilometre Array Phase 2 (SKA2), Wide Field Survey Telescope (WFST), Vera Rubin Observatory Legacy Survey of Space and Time (LSST), and Einstein Probe (EP), is presented.
We assume an exposure time of 1 hour for FAST and SKA2, 30 s for LSST and WFST, and 1500 s for Einstein probe.
As is shown in Fig. \ref{SIS}, galaxy-lensed orphan afterglows in the radio band and optical band could be detected.
The corresponding model parameters are: $z_{s}=1.0$, $z_{l}=0.5$, $M_{l}=10^{9}\ M_{\odot}$, $E=10^{52}$ erg, $\epsilon_{B}=0.01$, and $n=1.0\ \text{cm}^{-3}$.
We can estimate the peak flux of the ordinary afterglow from the equation~\citep[see review by][]{Zhang18},
\begin{equation}
F_{\nu, \max }=1.6 (1+z) \epsilon_{B,-2}^{1 / 2} E_{52} n^{-1} D_{L, 28}^{-2}~\mathrm{mJy}.
\end{equation}
Hereafter the convention $Q_{n}=Q / 10^{n}$ is adopted in c.g.s. units (e.g. $E_{52}=E /\left(10^{52} \mathrm{erg}\right)$ ).
It is found that the optical peak flux of the orphan afterglow is always lower than $F_{\nu, \max }$ by about one order of magnitude.
If the physical parameters of a jet satisfy the constraint equation,
\begin{equation}
(1+z) \epsilon_{B,-2}^{1 / 2} E_{52} n^{-1} D_{L, 28}^{-2}\geq \frac{F_{\text{LSST}}}{0.16~\mathrm{mJy}},
\end{equation}
where $F_{\text{LSST}}$ is the detection threshold (in \emph{r}-band) of LSST,
then its optical orphan afterglow can be observed by WFST and LSST in \emph{r}-band.
The emission in radio bands is also likely to be observed by SKA2.
Here, we take LSST as an example and estimate the detection rate of galaxy-lensed orphan afterglows.
The area of the whole celestial sphere $\Omega_{\text{sph}}$ is $41252.96\ \text{deg}^{2}$, and the field of view (FoV) for LSST $\Omega_{\text{FoV}}$ is $9.6\ \text{deg}^{2}$.
We assume that the operation time $t_{\text{ope}}$ for LSST is 8 hours per day, the exposure number of $n_{\text{exp}}$ for each visit is 1, the exposure time $t_{\text{exp}}$ is 30 s, and the other time of $t_{\text{oth}}$ for each visit is 15 s.
The duration $\Delta t$ that the brightness of the galaxy-lensed orphan afterglow is above the limiting magnitude of LSST, is about $0.8$ day.
The actual detection area of a sky survey is $\frac{\Omega_{\text{FoV}}t_{\text{ope}}}{n_{\exp } t_{\exp }+t_{\mathrm{oth}}}$, and the maximum detection probability of LSST for galaxy-lensed orphan afterglows is~\citep{Zhu21},
\begin{equation}
P_{\text{det}}=\frac{\Omega_{\text{FoV}}t_{\text{ope}}}{\Omega_{\text{sph}}\left(n_{\exp } t_{\exp }+t_{\mathrm{oth}}\right)}\Delta t.
\end{equation}
The resulting detection rate by LSST for galaxy-lensed orphan afterglows is $N \cdot P_{\text{det}}\lesssim0.08$ $\text{yr}^{-1}$.
Similarly, the detection rate by WFST is calculated, and the result is $\lesssim 0.02$ $\text{yr}^{-1}$.

On the other hand, the lensless orphan afterglows can not be detected by FAST or SKA2 in the radio band (see Fig. \ref{SIS}).
However, the peak flux of the galaxy-lensed orphan afterglow in the radio band is higher than the detection threshold of the FAST and LSST.
Similarly, the detection rates of SKA2 and FAST for galaxy-lensed orphan afterglows are estimated.
It is assumed that these two detectors perform the sky survey about a tenth of the operation time in a whole year, and their operation time is about 20 hours a day.
In order to detect orphan afterglow in the radio band, the exposure time of these two detectors needs to be an hour at least.
The area of the sky that these two detectors can monitor in a single sky survey is limited by the narrow field of view in the radio band and the long exposure time.
As a result, both SKA2 and FAST have a detection rate of $\lesssim5\times10^{-5}$ $\text{yr}^{-1}$.

In this section, the detection rate of galaxy-lensed orphan afterglows is calculated.
The optical emission of galaxy-lensed orphan afterglows can be detected by WFST and LSST in \emph{r}-band.
The peak flux of galaxy-lensed orphan afterglows in the radio band can be observed by SKA2 and FAST.
The radiation in the X-ray band of galaxy-lensed orphan afterglows is too faint to be detected by EP.
However, due to the narrow field of view and long exposure time, FAST and SKA2 have very low detection rates for galaxy-lensed orphan afterglows, which are less than $10^{-4}$ $\text{yr}^{-1}$.
It is found that the optical band is the best band to observe the galaxy-lensed orphan afterglows.

\vspace{-0.4cm}
\section{Conclusion and discussion}
\label{sec:summary}

In this paper,
the gravitationally lensed orphan afterglows of GRBs generated by structured jets are discussed in view of three different lens models, including the PM model, the SIS model, and the CR model.
A galaxy lens ($M_{l} \ge 10^{9}\ M_{\odot}$) under the SIS model could generate significant lensed signals in the optical/X-ray orphan afterglows,
in which the second peak appearing at the late stage ($10^{4} \sim 10^{5}$ s) could be
detected by survey telescopes in the \emph{r}-band, e.g., WFST and LSST.
The peak flux in the radio band of the galaxy-lensed orphan afterglow could be observed by SKA2 and FAST.
However, the X-ray emission of the galaxy-lensed orphan afterglow is below the threshold of EP.
It is found that the optical band is the best band to observe the galaxy-lensed orphan afterglows.

The lensed orphan afterglows of a Gaussian jet are also discussed (in Appendix \ref{sec:Gaussian}).
For the same $\theta_c$ and other jet parameters, the detection rate of the galaxy-lensed orphan afterglow corresponding to the Gaussian jet
will be slightly reduced than that of the power-law jet.
For these two possible jet structures, the event rate of galaxy-lensed orphan afterglows
is estimated to be $\lesssim$ 1.8 $\text{yr}^{-1}$ for the whole sky.
The optimistic detection rates of LSST and WFST for galaxy-lensed orphan afterglows are
$\lesssim$ 0.04$\sim$0.08 $\text{yr}^{-1}$ and $\lesssim$ 0.01$\sim$0.02 $\text{yr}^{-1}$ respectively.
On the other hand, the maximum detection rate is less than $10^{-4}$ $\text{yr}^{-1}$ for both SKA2 and FAST.
In addition, by analyzing the observed lensed orphan afterglow,
the redshift mass of the SIS lens could be derived with a relative error of $10\%$
through the autocorrelation function method when the time resolution is less than 330 s.

\citet{Mao92} showed that the probability of galaxy-lensed GRBs was between 0.05\% and 0.4\%.
Considering that the predicted GBM burst detection rate is about 200 bursts per year, the corresponding lensed event rate is 0.1$\sim$0.8 $\text{yr}^{-1}$.
The maximum of $\theta_{\text{v}}$ is limited to be $5^{\circ}$ in our calculations, and the event number of orphan afterglows is approximately 5 times the number of
on-axis afterglows.
The event rate of galaxy-lensed orphan afterglows corresponding to the result of \citet{Mao92} is 0.5$\sim$4.0 $\text{yr}^{-1}$, which is roughly consistent with our result of 1.8 $\text{yr}^{-1}$.
\citet{Paynter21} found a lensed GRB in a dataset of 2 679 bursts, so the rough estimation of the probability of strong gravitational lensing was $3.7_{-2.6}^{+7.8} \times 10^{-4}$ (90\% credibility).
The corresponding lensed event rate of orphan afterglows is roughly 0.4 $\text{yr}^{-1}$.
Using a GRB luminosity function and formation rate corrected for cosmological evolution, \citet{Hurley19} estimated the detection rate of lensed GRB pairs by Konus-Wind as
~0.02 $\text{yr}^{-1}$, which is in the same order of magnitude as our estimated detection rate of galaxy-lensed orphan afterglows by LSST.

It is still a challenge to detect and identify an orphan afterglow, although a large number of off-axis jet events are expected \citep{Ho18,Ho20b,Ho22}.
In practice, searching for orphan afterglows through a sky survey at the X-ray, optical, and radio band, one has to distinguish them from a lot of false or fake similar signals,
particularly from stellar flares at optical and X-ray wavelengths, and from the active galactic nucleus (AGN) at radio wavelengths \citep{Ho22}.
It further complicates the detection and identification that multiple data points must be obtained during the spike-like bump for identifying galaxy-lensed orphan afterglows.
Therefore, the actual detection rate of the LSST and WFST may be lower than our estimations.
However, some researchers have been devoted to methods to classify light curves of different transients from the observation data deluge,
e.g., machine learning techniques\citep{Hlozek20}.
With the development of these techniques, the success rate of identification of orphan afterglows is expected to be improved.

Some effects or events can produce a signal on orphan afterglows similar to the gravitational lensing effect of galaxies.
Most X-ray flares happen after the trigger hundreds to thousands of seconds, but some flares can be as late as $10^{6}$ s \citep{Falcone06,Margutti10},
which is comparable to the time scale of the bump generated by a galaxy lens with a mass of $10^{11}\ M_{\odot}$.
The radio emission of the orphan afterglow may be distorted due to the interstellar scintillation effect, producing frequency-dependent flux variations on timescales of minutes to days \citep{Alexander19}, which resemble the radio-band light curves of the galaxy-lensed afterglow.
The synchrotron or inverse Compton emission from the reverse shock can also induce additional spike-like bumps in the X-ray and radio light curves of orphan afterglows \citep{Wu03,Fan04,Geng18b}.
Multiple bumps caused by the gravitational lensing effect have quasi-identical spectra, and the time delay between them is frequency-independent.
The gravitational lensing effect can be distinguished from the X-ray flare, interstellar scintillation, and reverse shock emission
through detailed multi-band observations and spectral analysis.

The detection of galaxy-lensed orphan afterglows is of significant scientific value.
The gravitational lens effect could be used to probe the dark matter.
If an orphan afterglow is confirmed to be gravitationally lensed by a known galaxy, we can roughly obtain the mass proportion of dark matter in the galaxy by comparing the stellar mass of the galaxy with the estimated redshift mass of the galaxy.
The observation of galaxy-lensed orphan afterglows can also help to search for faint galaxies that are hard to be discovered by optical instruments.
With the development of detection technology, the orphan afterglow database may be a good reservoir for searching lensed events, especially in the coming era of the time-domain survey.

\vspace{-0.4cm}
\section*{Acknowledgements}
We appreciate the anonymous referee for constructive suggestions.
This work is supported by the National Natural Science Foundation of China (Grant Nos. 11903019, 11873030, 12041306, 11833003, U1938201, 11725314, 11535005),
by the National SKA Program of China No. 2020SKA0120300,
by the National Key R\&D Program of China (2021YFA0718500), by the Major Science and Technology Project of Qinghai Province (2019-ZJ-A10),
and by the science research grants from the China Manned Space Project with NO. CMS-CSST-2021-B11.

\vspace{-0.4cm}

\section*{Data availability}
No new data were generated or analysed in support of this research.

\bibliographystyle{mnras}
\bibliography{export-bibtex} 
\vspace{-0.5cm}

\bsp	
\label{lastpage}

\appendix
\section{Gaussian structured jet}
\label{sec:Gaussian}
Here, we discuss the gravitationally lensed orphan afterglows produced by a Gaussian structured jet.
A Gaussian structured jet is assumed as
\begin{equation}
\begin{gathered}
\varepsilon(\theta)=\varepsilon_{c}  \exp \left(-\frac{1}{2} \frac{\theta^{2}}{\theta_{c}^{2}}\right),\\
\Gamma(\theta)=\Gamma_{c}  \exp \left(-\frac{1}{2} \frac{\theta^{2}}{\theta_{c}^{2}}\right)+1.
\end{gathered}
\end{equation}

\begin{figure*}
\vskip-0.1in
\includegraphics[angle=0,scale=0.55]{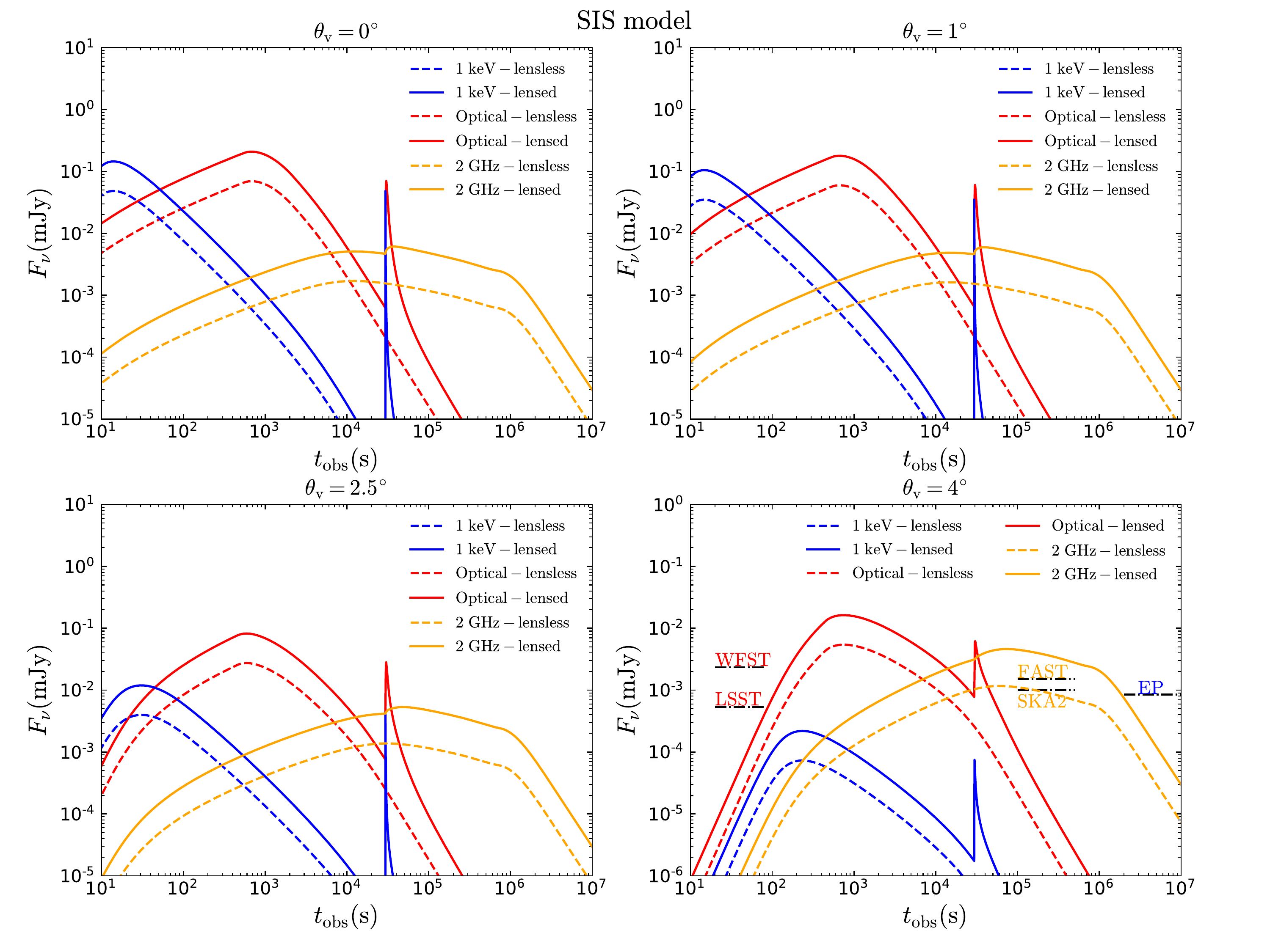}
    \caption{Lensed afterglow light curves produced
    by a Gaussian structured jet for the SIS model.
    Line styles are the same as Fig. \ref{PM}.
    }
    \label{SISG}
\end{figure*}

Parameters such as $\theta_{c}$  are the same as in Sec.  \ref{sec:lensmodel},
and the lensed afterglow light curves corresponding to the Gaussian structured jet are calculated.
It is found that the shapes of light curves generated by the Gaussian structured jet are similar to those from the power-law structured jet.
For example, the SIS model lensed afterglows of the Gaussian structured jet are exhibited in Fig. \ref{SISG}.
As is shown, a significant bump exists for large $\theta_{v}$ in both the X-ray and optical band.
However, since the kinetic energy density $\varepsilon(\theta)$ of the Gaussian structured jet decays faster with the jet angle $\theta$,
for $\theta_{\text{v}}>\theta_{c}$, the orphan afterglow of the Gaussian jet is fainter than that of the power-law jet at the same viewing angle.
Therefore, although the peak flux of the on-axis afterglow in Fig. \ref{SISG} is consistent with that in Fig. \ref{SIS},
the corresponding upper limit of viewing angle for orphan afterglows is $4^{\circ}$ and smaller than that in Fig. \ref{SIS}.
Similar conclusions could be derived for cases of the other two gravitational lens models.

If we assume a typical value of $\theta_{c}$ ($=2^{\circ}$) for these two possible structured jets, the detection rate of the galaxy-lensed orphan afterglow corresponding to the Gaussian jet would be half of those for the power-law jet.
The true structure of the GRB jet is still under debate, and both the power-law jet and Gaussian jet are possible.
At the same time, there is a strong degeneracy among the parameters in the GRB afterglow model.
The afterglow light curves of the two distinct structured jets can have little difference although their $\theta_{c}$ are different.

\end{document}